\documentclass[twocolumn,nofootinbib]{nature}
\usepackage[colorlinks]{hyperref}
\usepackage{amsmath,mathrsfs,amsbsy,color,graphicx,bm,amsthm,amsfonts,dsfont}
\hypersetup{
    colorlinks = true,
    citecolor = {blue},
    linkcolor = {red}
}
\usepackage{epstopdf}
\usepackage{indentfirst}
\usepackage{graphicx}
\usepackage{amsmath}
\usepackage{latexsym}
\usepackage{color}
\usepackage{bbm}
\usepackage{upgreek}

\begin{document}
\title{Distributed quantum phase estimation with entangled photons}

\author{Li-Zheng Liu$^{1,2,3,*}$, Yu-Zhe Zhang$^{1,2,3,*}$, Zheng-Da Li$^{1,2,3}$, Rui Zhang$^{1,2,3}$, Xu-Fei Yin$^{1,2,3}$ Yue-Yang Fei$^{1,2,3}$, Li Li$^{1,2,3}$, Nai-Le Liu$^{1,2,3}$, Feihu Xu$^{1,2,3}$, Yu-Ao Chen$^{1,2,3}$ \& Jian-Wei Pan$^{1,2,3}$}
\maketitle

\begin{affiliations}
\item Hefei National Laboratory for Physical Sciences at the Microscale and Department of Modern Physics, University of Science and Technology of China, Hefei 230026, China
\item Shanghai Branch, CAS Center for Excellence in Quantum Information and Quantum Physics, University of Science and Technology of China, Shanghai 201315, China
\item Shanghai Research Center for Quantum Sciences, Shanghai 201315, China     \\
$^{*}$These authors contribute equally.
\end{affiliations}

\maketitle

\begin{abstract}
Distributed quantum metrology can enhance the sensitivity for sensing spatially distributed parameters beyond the classical limits. Here we demonstrate distributed quantum phase estimation with discrete variables to achieve Heisenberg limit phase measurements. Based on parallel entanglement in modes and particles, we demonstrate distributed quantum sensing for both individual phase shifts and an averaged phase shift, with an error reduction up to 1.4 dB and 2.7 dB below the shot-noise limit. Furthermore, we demonstrate a combined strategy with parallel mode entanglement and multiple passes of the phase shifter in each mode. In particular, our experiment uses six entangled photons with each photon passing the phase shifter up to six times, and achieves a total number of photon passes $N=21$ at an error reduction up to 4.7 dB below the shot-noise limit. Our research provides a faithful verification of the benefit of entanglement and coherence for distributed quantum sensing in general quantum networks.
\end{abstract}

\emph{Introduction. ---}
Quantum metrology exploits the quantum mechanical effects to increase the sensitivity of precision sensors beyond the classical limit\cite{giovannetti2004quantum,giovannetti2011advances,degen2017quantum}. By using the entanglement or coherence of the quantum resource\cite{braun2018quantum}, it can achieve higher precision of parameter estimation below the shot-noise limit (SNL), and its sensitivity can saturate the Heisenberg limit\cite{kok2002creation,walther2004broglie,mitchell2004super,nagata2007beating,resch2007time,higgins2007entanglement,gao2010experimental,bell2013multicolor,slussarenko2017unconditional}, which is believed to be the maximum sensitivity achievable over all kinds of probe quantum state.

It is well known that many important applications can be regarded as sensor networks with spatially distributed parameters, which is often referred to distributed quantum metrology\cite{schnabel2010quantum,aasi2013enhanced,humphreys2013quantum,perez2012fundamental}. In the framework of distributed quantum metrology, an important class of estimation problems is concerned with the sensing of individual parameters. Typically, the Heisenberg limit can be achieved in both continuous-variable and discrete-variable states\cite{polino2020photonic}. However, recently, there has been increasing interest in the study of multiparameter estimation, particularly in the linear combination of the results of multiple simultaneous measurements at different locations (or modes), for example, averaged phase shift. For instance, estimating the averaged phase shift among remote modes is the fundamental building block to construct a quantum-enhanced international timescale (world clock)\cite{komar2014quantum}; in classical sensing, averaged phase shift is widely used to evaluate the gas concentration of a hazardous gas in a given area\cite{1668063,2008Geographic}, where the entangled network can substantially enhance the sensing accuracy. In such a case, estimating the parameters separately in modes is not optimal. Even with particle entanglement in each mode, the root-mean-square error (RMSE) for the estimation of the linear combination of multiple parameters is restricted to $\sqrt M/N$, where $N$ denotes the total number of entangled particles among all $M$ modes with $M\leq N$. However, the ultimate Heisenberg limit is $~1/N$. In contrast, the entanglement among modes in an entangled network can substantially enhance the sensitivity for multiparameter estimation\cite{zhuang2018distributed,slussarenko2017unconditional}. Remarkable experiments have demonstrated distributed quantum metrology with multi-mode entangled continuous-variable state\cite{guo2019distributed,PhysRevLett.124.150502}.

Recently, it has been shown that if the distributed sensors are entangled in both modes and particles, it is possible to achieve the ultimate Heisenberg limit\cite{ge2018distributed,proctor2018multiparameter,gessner2018sensitivity,oh2020optimal}. Here, we refer to entanglement in both modes and particles as the parallel strategy. Furthermore, it has been shown that the sequential scheme --- a single probe interacting coherently multiple times with the sample --- can be used to reach the same Heisenberg limit~\cite{higgins2007entanglement,braun2018quantum}, and this has been shown to be an optimal parameter estimation strategy in several applications\cite{juffmann2016multi,hou2019control}. Also, the combination of sequential scheme and parallel entanglement can even outperform parallel strategy for the estimation of multiple parameters in the presence of noise\cite{higgins2007entanglement,xiang2011entanglement,berni2015ab,daryanoosh2018experimental}. We refer this combination as the combined strategy.

In this Article, we perform the experimental demonstration of discrete-variable distributed quantum metrology for both individual phase shifts and averaged phase shift. In the parallel strategy for estimating individual phase shifts, by preparing three high-fidelity two-photon entangled sources, we demonstrate three individual phase-shift measurements, where the distributed sensors of (mode 1, mode 2, mode 3) achieve a super-resolution effect with RMSE reductions up to (1.44 dB, 1.43 dB, 1.43 dB). In the parallel strategy for estimating averaged phase shifts, by constructing a high-fidelity multiphoton interferometer, we compare the sensitivities for the scenarios of modes entangled/separated and particles entangled/separated, which are referred to as MePe, MePs, MsPe, MsPs, respectively. The results show that compared with the SNL of MsPs, the distributed sensors of (MePe, MePs, MsPe) can achieve a precision of RMSE reduction up to (2.7 dB, 1.56 dB, 1.43 dB) for the estimation of an averaged phase shift across three modes. In the combined strategy, by interacting the photons with the phase shifter multiple times in each mode, we perform a demonstration for the estimation of an unequal-weight linear function of multiple phase shifts across six entangled modes. The experiment realizes a total number of photon-passes at $n=21$ with an error reduction up to 4.7 dB below the SNL. Note that the evaluation of the super-resolution effect in these experiments uses post selection that does not include the photon losses\cite{slussarenko2017unconditional}.

\begin{figure}[htbp!]
	\centering
	\includegraphics[width=0.6\linewidth]{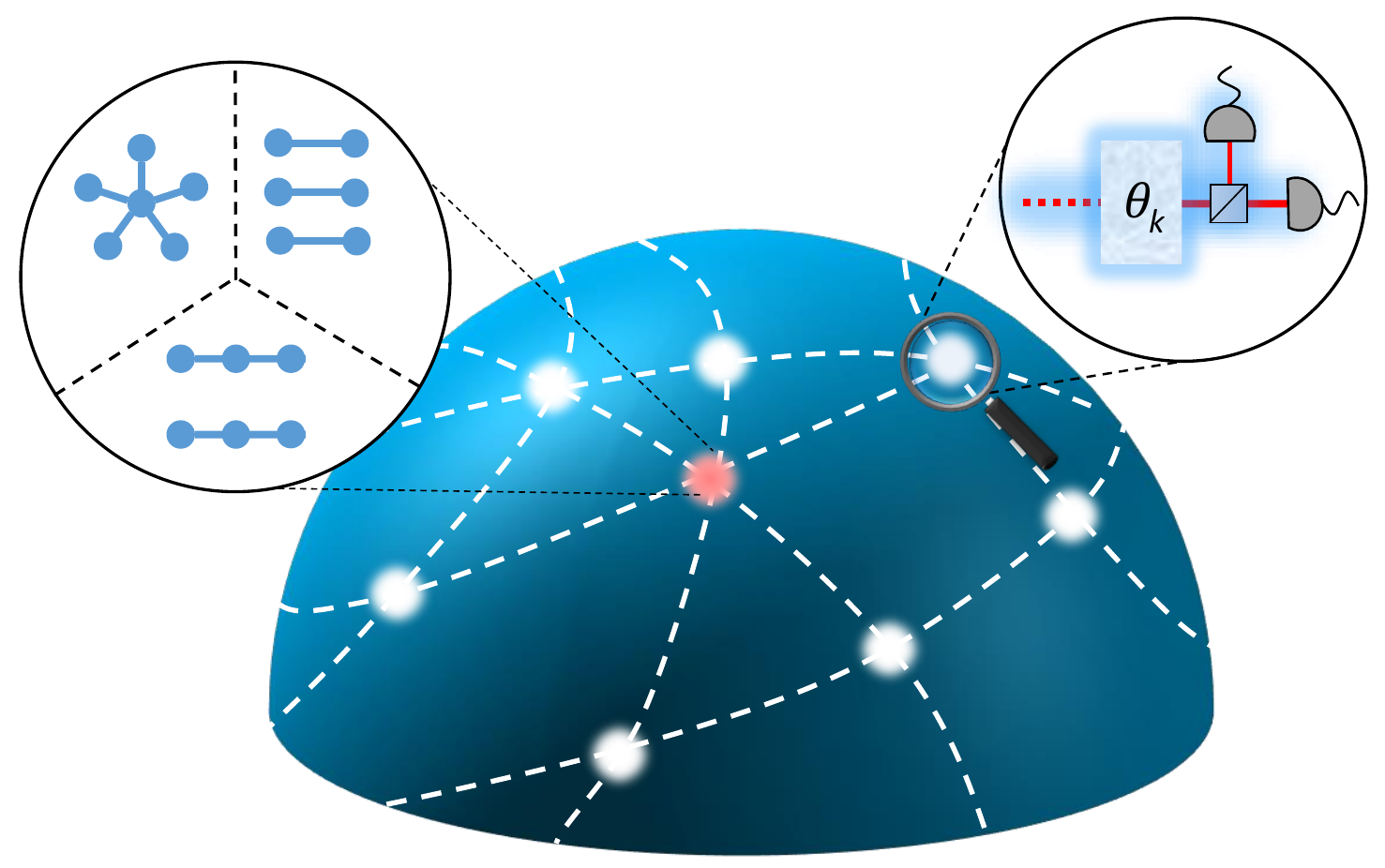}
	\caption{\textbf{A sketch for estimating distributed multiparameter.} Each sensor node (white) is equipped with a measurement device where the red dashed line represents a single pass or multiple passes and $\theta_k$ represents the unknown phase shift in mode $k$, while the central node (red) is equipped with a source to produce multi-party entangled states. The white dashed lines represent the quantum channel that can be used to distribute photons.}
	\label{fig:randomness4}
\end{figure}

\emph{Protocol. ---}
Let us consider a general scenario in the framework of distributed quantum metrology with $\emph{M}$ modes. As shown in Fig.~\ref{fig:randomness4}, we assume that each mode has an unknown phase shift $\theta_k$. In the estimation of individual parameters, we assume the quantity to be estimated is three individual phase shifts. The probe states in our experiment have the form $ \left| \phi_{\rm individual} \right\rangle = \frac{1}{\sqrt{2}}(|HH\rangle + |VV\rangle)$ for each mode, where $H (V)$ denotes the horizontal (vertical) polarization. In the multiparameter strategy, we assume that the quantity to be estimated is a linear global function $\hat{\theta} =  \bm{\upalpha^\text{T}}\bm{\uptheta}$, where ${\bm{\uptheta}} = ({\theta}_1,\dots,{\theta}_M)$ and $ \bm{\upalpha}=(\alpha_1,\dots,\alpha_M)$ denote, respectively, the vector of phase shift and the normalization coefficients for the mode $k = 1,\dots,M$. $\bm{\upalpha^\text{T}}$ is the transpose of $\bm{\upalpha}$. The unitary evolution is given by,
\begin{equation}
\hat{U}(\bm{\uptheta}) = e^{-i\sum^{M}_{k=1}\hat{H}_k\theta_k}= e^{-i\bm{\hat{\text{H}}}\cdot\bm{\uptheta}},
\end{equation}
where $ \bm{\hat{\text{H}}} = (\hat{H}_1,\dots,\hat{H}_M) $ are the Hamiltonians governing the evolution. The task is to estimate $\hat{\theta}$ with a high precision using classical or quantum sensor networks. Here the uncertainty (or error) for the estimation of $\hat{\theta}$ can be described by $ \Delta \hat{\theta}=\left(\bm{\upalpha}^{\mathrm{T}}\Sigma\bm{\alpha} \right)^{1/2}$ with multiparameter covariance matrix $\Sigma$ whose elements are $\Sigma_{km} = E[(\theta_k-\theta_{est,k})(\theta_m-\theta_{est,m})]$, where $\theta_{est,k}$ and $\theta_{est,m}$ denote, respectively, the locally unbiased estimator for $\theta_k$ and $\theta_m$ for $k,m = 1,\dots,M$. $E[X]$ denotes the expectation value of the random variable $X$.

(1) In the parallel strategy, we set three probe modes where the objective function to be estimated is the averaged phase shift $\hat{\theta}=\sum_{k=1}^{3}{\theta_k}/3$, and the Hamiltonians are set to $ \hat{H}_k = \sigma_z/2 $ for mode $k =1,2,3$, where $\sigma_z$ is the Pauli matrix. Thus the evolution can be described as
\begin{equation} \label{Uk}
\hat{U}_{k} =
\begin{pmatrix}
e^{-i\theta_k/2} & 0 \\
0 & e^{i\theta_k/2}
\end{pmatrix}.
\end{equation}
The overall probe states in this scheme can be classified as the modes entangled and particles entangled (MePe), modes entangled and particles separated (MePs) and modes separated and particles entangled (MsPe), which have the form of,
\begin{align} \label{eq5}
&|\phi_{\text{MePe}}\rangle=\frac{1}{\sqrt{2}}(|HH\rangle_{M_1}|HH\rangle_{M_2}|HH\rangle_{M_3}+|VV\rangle_{M_1}|VV\rangle_{M_2}|VV\rangle_{M_3}),\\ \nonumber
&|\phi_{\text{MePs}}\rangle=\frac{1}{2}\left (|H\rangle_{M_1}|H\rangle_{M_2}|H\rangle_{M_3}+|V\rangle_{M_1}|V\rangle_{M_2}|V\rangle_{M_3}\right )^{\otimes2},\\ \nonumber
&|\phi_{\text{MsPe}}\rangle=\frac{1}{2\sqrt{2}}\left (|HH\rangle+|VV\rangle\right )_{M_1}\left (|HH\rangle+|VV\rangle\right )_{M_2}\left (|HH\rangle+|VV\rangle\right )_{M_3}.
\end{align}

(2) In the combined strategy, beside the parallel entanglement among modes, we utilize the coherence rather than the particle entanglement in each mode. In this case, the essential feature is that the phase shift is being interacted coherently over many passes of the unitary evolution, $\hat{U}_{k}$. We implement a distributed phase sensing with six modes, and the objective functions to be estimated are unequal weighted linear functions: $\hat{\theta}=\sum_{k=1}^{6}{k \theta_k}/21$. For each modes $k$, the evolution with multiple passes $j$ are set to,
\begin{equation}
\hat{U}_{k}^{j}=\prod_{i=1}^{j} {\hat{U}_{k}},
\end{equation}
where $\hat{U}_{k}$ is same as Eq.~\eqref{Uk}. To demonstrate this protocol, we define two types of probe states, namely modes entangled and particles coherent (MePc) and modes separated and particles coherent (MsPc), which can be written as
\begin{align} \label{eq6p}
&|\phi_{\text{MePc}}\rangle=\frac{1}{\sqrt{2}}(|H\rangle_{M_1}|H\rangle_{M_2}|H\rangle_{M_3}|H\rangle_{M_4}|H\rangle_{M_5}|H\rangle_{M_6}+|V\rangle_{M_1}|V\rangle_{M_2}|V\rangle_{M_3}|V\rangle_{M_4}|V\rangle_{M_5}|V\rangle_{M_6}),\\ \nonumber
&|\phi_{\text{MsPc}}\rangle=\otimes_{Mi=M_1}^{M_6}\left( \frac{1}{\sqrt{2}}\left (|H\rangle+|V\rangle\right )_{M_i} \right ).
\end{align}

The projective measurements on the probe states are performed in the $ \sigma_x $ basis, which can achieve the maximum visibility for interference fringe\cite{gessner2018sensitivity,ge2018distributed}. In this setting, the outcome probability in the eigenvectors $|\pm1\rangle$ can be written as
\begin{equation} \label{eq4}
\ P_{\pm1} =\frac{1\pm V_{\pm}\cos \hat{\theta} }{2},
\end{equation}
where $V_\pm$ is the interference fringe visibility for $n$-Greenberger-Horne-Zeilinger states\cite{gao2010experimental,resch2007time}. The widely adopted elementary bounds on the RMSE are given by the Cramer-Rao bound $\Delta \hat{\theta} \geq \frac{\bm{\alpha^T}\bm{\alpha}}{\sqrt{\mu \bm{\alpha^T}\bm{\text{F}}\bm{\alpha}}}$, where $\mu$ denotes the number of independent measurements and $\bm{\text{F}}$ denotes the classical Fisher matrix with elements $(\bm{\text{F}})_{kl}=\sum\nolimits_{i=\pm1}P_{i}\left[(\partial/\partial\theta_{k})P_{i} \right]\left[(\partial/\partial\theta_{l})P_{i} \right]$. The effective Fisher information (FI)\cite{helstrom1969quantum}, $F(\hat{\theta})$, can be used to evaluate the estimation sensitivity\cite{giovannetti2011advances}, and it is given by,
\begin{equation} \label{eq:FI}
F(\hat{\theta})=\frac{ \bm{\upalpha^T}\bm{\text{F}}\bm{\upalpha}}{(\bm{\upalpha^T}\bm{\upalpha})^2}.
\end{equation}
By combining Eq.~\eqref{eq4} with equation Eq.~\eqref{eq:FI}, we can calculate the FI for the linear function $\hat{\theta}$. Note that the FI is used to quantify that the accuracy for different strategies (Methods) and the calculations of FI use only the post selected photons, which does not include photon losses.

\begin{figure*}
	\centering
	\includegraphics[width=\linewidth]{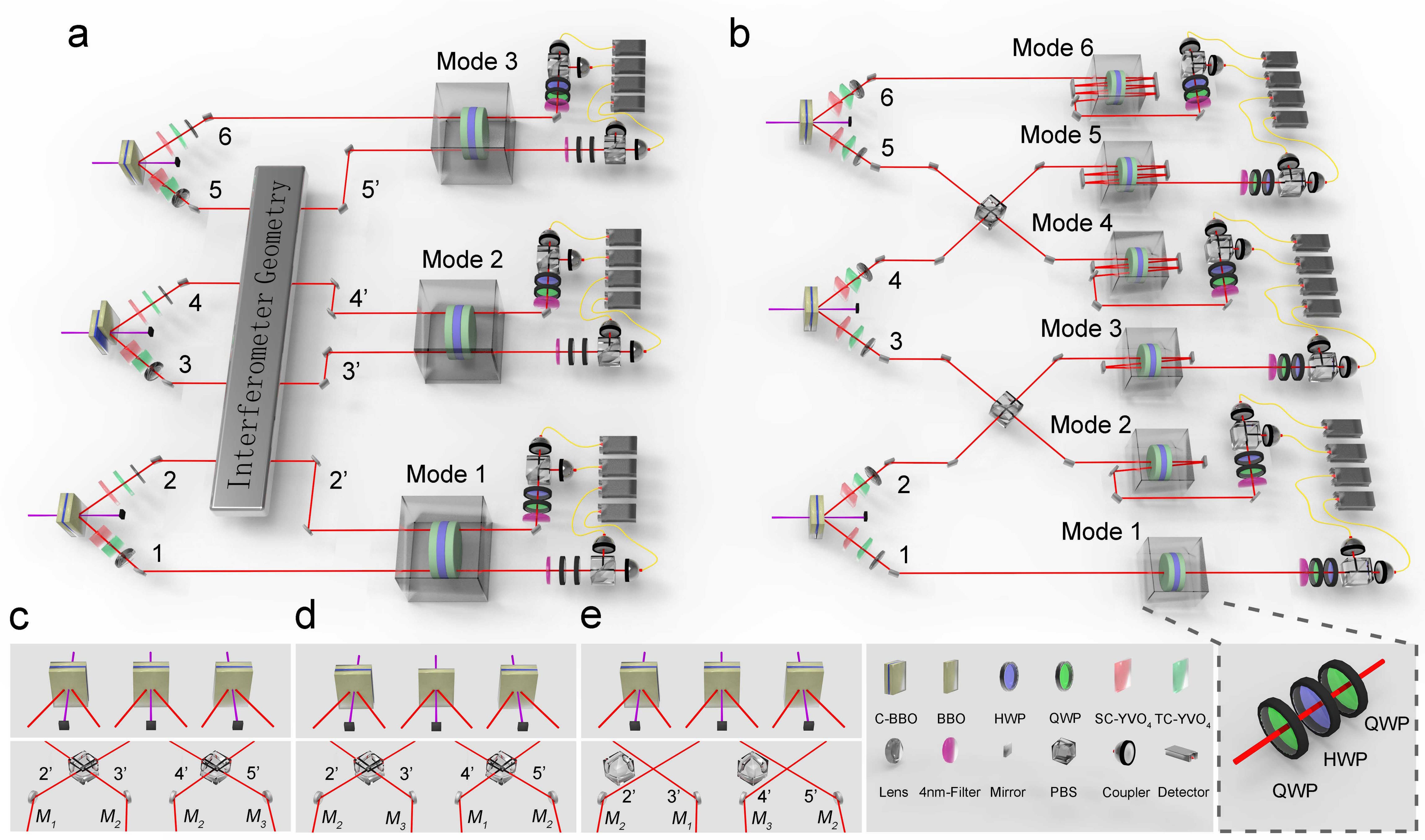}
	\caption{\textbf{Experimental set-up.} \textbf{a,b,} The set-up for the parallel strategy (\textbf{a}) and the combined strategy (\textbf{b}). An ultrafast ultraviolet pump laser passes BBO(s) or C-BBO(s) to produce three down-converted photon pairs. In each channel, a lens is used to ensure the collimation of the beam. The narrow-bandpass filters with full-width at half-maximum (FWHM) wavelengths of $\lambda_{\rm FWHM}$ = 4 nm are used to suppress frequency-correlated effect between the signal photon and the idler photon. The probes prepared by different combinations of the interferometer, are distributed to three modes and then undergo the evolution. Finally, the probes are detected by measurement systems, consisting of PBS, HWP, QWP and two single-photon detectors. A complete set of $2^6$ six-photon coincidence events are simultaneously registered. \textbf{c-e} The interferometer configurations to produce the quantum states of MePe(\textbf{c}), MePs(\textbf{d}), MsPe(\textbf{e}). SC-YVO$_4$ and TC-YVO$_4$ represent spatial compensation (SC) and temporal compensation (TC) yttrium orthovanadate crystals (YVO$_4$)}
	\label{fig:setup}
\end{figure*}

\emph{Experimental set-up. ---}
The experimental set-up is illustrated in Fig.~\ref{fig:setup}a,b. A pulsed ultraviolet laser---with a central wavelength of 390 nm, a pulse duration of 150 fs and a repetition rate of 80 MHz---is focused on three sandwich-like combinations of BBO crystals (C-BBO) to generate the independent entangled photon pairs in the channel 1 and 2, 3 and 4, and 5 and 6. With this configuration, the photon pairs are generated in the state $ \left| \phi^{+} \right\rangle = \frac{1}{\sqrt{2}}(|HH\rangle + |VV\rangle)$, where $H (V)$ denotes the horizontal (vertical) polarization.

In the parallel strategy (Fig.~\ref{fig:setup}a), the initial probe states $\left| \phi_{\rm MePe} \right\rangle$, $\left| \phi_{\rm MePs} \right\rangle$, $\left| \phi_{\rm MsPe} \right\rangle$ can be prepared by combining three independent spontaneous parametric down conversion (SPDC) sources and a tunable interferometer with the platforms as shown in Fig.~\ref{fig:setup}c-e. In the combined strategy (Fig.~\ref{fig:setup}b), the photon in mode $k$ coherently passes through the phase shift $k$ times, and Hong-Ou-Mandel (HOM) interferences between photons 2 and 3 and photons 4 and 5 are applied. In each channel, a lens is used to ensure the collimation of the beam. The narrow-bandpass filters with full-width at half-maximum (FWHM) wavelengths of $\lambda_{\rm FWHM}$ = 4nm are used to suppress frequency-correlated effect between the signal photon and the idler photon. The probe states evolve and pass through quatre wave plate (QWP) and half wave plate (HWP), and finally detected by single-photon detectors.

The detailed configurations of the tunable interferometer with four inputs and four outputs are shown in Fig.~\ref{fig:setup}c-e. The tunable interferometer consists of two polarizing beam splitters ($\rm PBS_{1}$ and $\rm PBS_{2}$) controlled by multi-axis translation stages, whose position at left and right corresponds to non-interference and interference between photon 2 and 3 (4 and 5). The three C-BBOs generate three Einstein-Podolsky-Rosen entangled photon pairs in the states $\left| \phi^{+}_{12} \right\rangle$, $\left| \phi^{+}_{34} \right\rangle$, $\left| \phi^{+}_{56} \right\rangle$. As shown in Fig.~\ref{fig:setup}c, to prepare the state $\left| \phi_{\rm MePe} \right\rangle$, the positions of both $\rm PBS_{1}$ and $\rm PBS_{2}$ controlled by two multi-axis translation stages are set to the right; the two PBSs will introduce the Hong-Ou-Mandel interference between photons 2 and 3 and photons 4 and 5. By replacing one of the C-BBOs with a single piece of BBO crystal as shown in Fig.~\ref{fig:setup}d, the state of $\left| \phi_{\rm MePs} \right\rangle$ is produced. We obtain the down-conversion probability of the prepared uncorrelated state is about $p=0.0110$. In Fig.~\ref{fig:setup}e, the positions of both $\rm PBS_{1}$ and $\rm PBS_{2}$ are set to the left, where there is no interference; this leads to the prepared quantum state of $\left| \phi_{\rm MsPe} \right\rangle$ and $\left| \phi_{\rm individual} \right\rangle$. They have a typical down-conversion probability of $p=0.0195 $ per pulse and a fidelity 0.9866 $\pm$ 0.0002.

\begin{figure*}
	\centering
	\includegraphics[width=\linewidth]{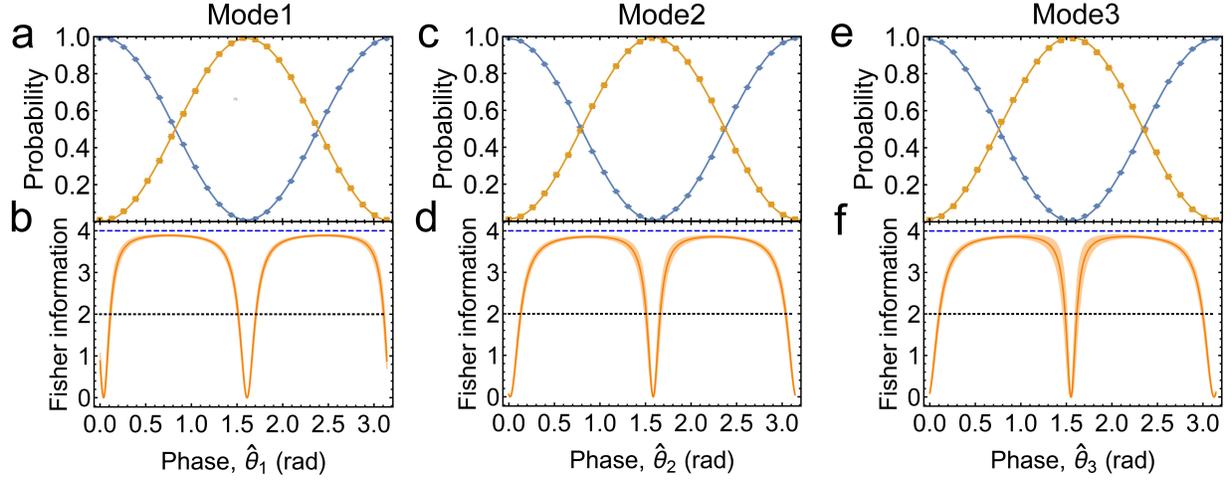}
	\caption{\textbf{Experimental results for estimating single parameters for mode 1, mode 2 and mode 3.} \textbf{a,c,e,} The average outcome probability for mode 1 (\textbf{a}), mode 2 (\textbf{c}) and mode 3 (\textbf{e}) in the measurement basis $\sigma_x^{\otimes2}$ for two-photon entangled states. The blue (orange) lines represent the average outcome probability $P_{+}^{13^{\prime}}$, $P_{+}^{2^{\prime}5^{\prime}}$ and $P_{+}^{4^{\prime}6}$ ($P_{-}^{13^{\prime}}$, $P_{-}^{2^{\prime}5^{\prime}}$ and $P_{-}^{4^{\prime}6}$) for mode 1, mode 2 and mode 3. \textbf{b,d,f,} The FI per trial, fitted from $\{P^{13^{\prime}}, P^{2^{\prime}5^{\prime}}, P^{4^{\prime}6}\}$ for mode 1 (\textbf{b}), mode 2 (\textbf{d}) and mode 3 (\textbf{f}). The shaded areas correspond to the $90\%$ confidence region, which are obtained from the uncertainty in the fitting parameters. Error bars are calculated from measurement statistics and too small to be visible. The blue dashed line is the theoretical limit of FI for the Heisenberg limit. The black dotted line is the theoretical limit of FI for the SNL.}
	\label{fig:randomness1}
\end{figure*}

\emph{Results. ---}
In the estimation of individual phases, each mode occupies a two-photon entangled state. We direct the photon 1 and $3^{\prime}$, $2^{\prime}$ and $5^{\prime}$, and $4^{\prime}$ and $6^{\prime}$ to mode 1, mode 2 and mode 3 respectively, and introduce the phase shifts with $\{\theta_1,\ \theta_2,\ \theta_3\}$ continuously from $ 0 $ to $ \pi$. The values to be estimated are three individual phase shifts $\{\theta_1,\ \theta_2,\ \theta_3\}$. The results are shown in Fig.~\ref{fig:randomness1}. We obtain the visibility $\{0.982,\ 0.989\}_{13^{\prime}}$, $\{0.989,\ 0.974\}_{2^{\prime}5^{\prime}}$ and $\{0.992,\ 0.970\}_{4^{\prime}6}$ for mode $k=1, 2$ and 3, and the optimal FIs are 3.88, 3.85 and 3.86, respectively. This also forms the results for MsPe.

\begin{figure*}
	\centering
	\includegraphics[width=\linewidth]{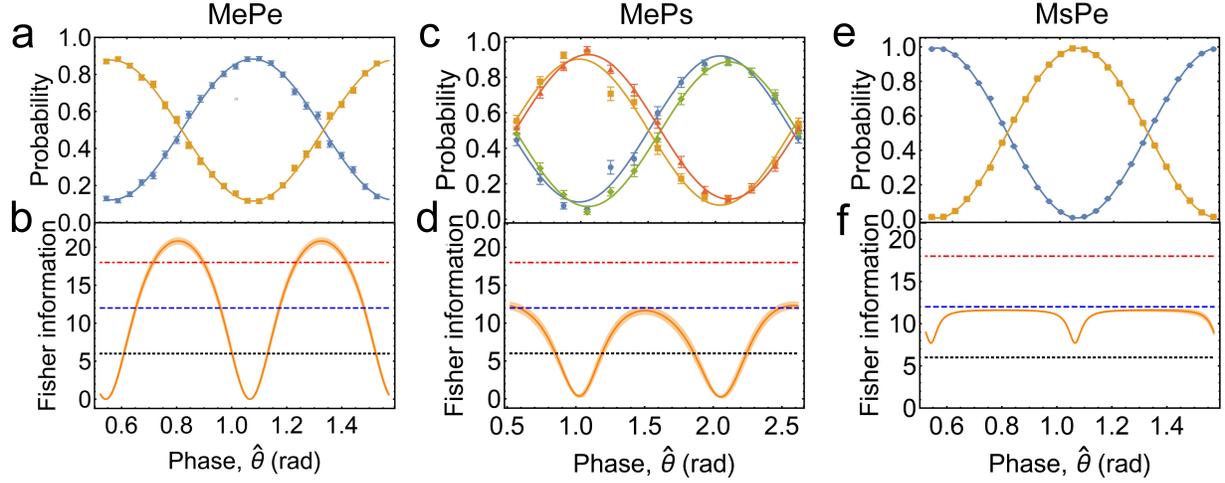}
	\caption{\textbf{Experimental results for the parallel strategies of MePe, MePs and MsPe.} \textbf{a,c,e,} The average outcome probabilities in the measurement basis $\sigma_x^{\otimes6}$, $\sigma_x^{\otimes3}$, $\sigma_x^{\otimes2}$ for six-photon (\textbf{a}), three-photon (\textbf{c}) and two-photon (\textbf{e}) entangled states. In \textbf{a}, the blue (orange) line represents the average outcome probability $P_{+}^{12^{\prime}3^{\prime}4^{\prime}5^{\prime}6}$ ($P_{-}^{12^{\prime}3^{\prime}4^{\prime}5^{\prime}6}$). In \textbf{c}, the blue (orange) line represents the average outcome probability $P_{+}^{12^{\prime}3^{\prime}}$ ($P_{-}^{12^{\prime}3^{\prime}}$) and the green (red) line represents the average outcome probability $P_{+}^{4^{\prime}5^{\prime}6}$ ($P_{-}^{4^{\prime}5^{\prime}6}$). In \textbf{e}, the blue (orange) line represents the average outcome probability $P_{+}^{13^{\prime}}$ ($P_{-}^{13^{\prime}}$). \textbf{b,d,f,} The FI per trial, fitted from $P^{12^{\prime}3^{\prime}4^{\prime}5^{\prime}6}$, $\{P^{12^{\prime}3^{\prime}}$, $P^{4^{\prime}5^{\prime}6}\}$ and $\{P^{13^{\prime}}$, $P^{2^{\prime}5^{\prime}}$, $P^{4^{\prime}6}\}$ for MePe (\textbf{b}), MePs (\textbf{d}) and MsPe (\textbf{f}). The shaded areas correspond to the 90$\%$ confidence region, which are obtained from the uncertainty in the fitting parameters. Error bars are calculated from measurement statistics. The red dot-dashed line is the theoretical limit of FI for MePs. The blue dashed line is the theoretical limit of FI for MsPe. The black dotted line is the theoretical value of FI for MsPs.}
	\label{fig:randomness2}
\end{figure*}

In the parallel strategy for estimating averaged phase shift, after the evolutions by QWPs and HWP, the coincidence measurements in the basis $ \sigma_x^{\otimes 6}$ are performed. Fig.~\ref{fig:randomness2}a shows the observed average outcome probability values with $\theta_1$ ramping continuously from $ 0 $ to $ \pi $ where we fix $\theta_2 = \frac{\pi}{6}, \theta_3 = \frac{\pi}{3} $. The experimental data are fitted to the function $\ P_{\pm}=\frac{1\pm V_\pm \cos(6\hat{\theta})}{2}$, where $V_{+}=0.756$ and $V_{-}=0.765$ denote the fringe visibilities in the eigenvectors $|\pm1\rangle$ of measurement basis ${\sigma_x}^{\otimes6}$. To quantify the sensitivity, we calculate the FI according to Eq.~\eqref{eq:FI}. As shown in Fig.~\ref{fig:randomness2}b, we demonstrate an enhancement of sensitivity for a range of phase shifts, and the maximum value of FI is about 20.825 at $\hat{\theta}=\frac{\pi}{6}$, which represents a 2.70-dB reduction as compared with the SNL of MsPs.

For the protocol with MePs, the photons 1 and $4^{\prime}$, $2^{\prime}$ and $5^{\prime}$, and $3^{\prime}$ and 6 are directed to mode 1, mode 2 and mode 3, respectively. Similar to MePe, we fix the phase shift $\theta_2 = \frac{\pi}{6},\ \theta_3 = \frac{\pi}{3}$ and change the phase shift $ \theta_1 $ continuously from $ 0 $ to $ 2\pi $. For photons 1, $2^{\prime}$ and $3^{\prime}$ ($4^{\prime}$, $5^{\prime}$ and 6), the outcome probability values can be obtained by observing the counts of photons $4^{\prime}$, $5^{\prime}$ and 6 (1, $2^{\prime}$ and $3^{\prime}$). Fig.~\ref{fig:randomness2}c shows the observed outcome probability values. In this case, the fitting function is set to $P_{\pm}^{12^{\prime}3^{\prime}}=P_{\pm}^{4^{\prime}5^{\prime}6}=\frac{1\pm V_\pm \cos ( 3\hat{\theta} )}{2}$. We obtain the fringe visibility in the eigenvectors $|\lambda=\pm1\rangle$ of measurement basis ${\sigma_x}^{\otimes3}$ as $\{0.808, 0.838\}_{12^{\prime}3^{\prime}}$ and $\{0.885, 0.878\}_{4^{\prime}5^{\prime}6}$. From Fig.~\ref{fig:randomness2}d, we also demonstrate the sensitivity enhancement, and the maximum value of FI is 12.313 at the phase shift $\hat{\theta}=\pi/6$, which is a 1.56-dB reduction over the SNL of MsPs.

With a little difference, the value to be estimated in the strategy of multiparameter estimation is the linear function $\hat{\theta}=\frac{1}{3}(\theta_1+\theta_2+\theta_3)$. The fitting function is set to $\ P^{13^{\prime}}_{\pm}=P^{2^{\prime}5^{\prime}}_{\pm}=P^{4^{\prime}6}_{\pm}=\frac{1\pm V_\pm \cos\left ( 2\theta_k \right )}{2}$ for mode $k=1,2$ and 3. The results of the protocol with MsPe are shown in Fig.~\ref{fig:randomness2}e,f. We obtain the visibility $\{0.985,\ 0.986\}_{1\&3'}$, $\{0.979,\ 0.979\}_{2'\&5'}$ and $\{0.994,\ 0.976\}_{4'\&6}$ for mode $k$=1, 2 and 3, respectively. The optimal FI of mode 1 is about 3.887, and the optimal FIs of mode 2 and mode 3 are 3.832 and 3.877, respectively (Extended Data Figure 1). The results in Fig.~\ref{fig:randomness2} clearly demonstrate that MePe is the best choice with the highest sensitivity to estimate the global function $\hat{\theta}$.

\begin{figure}
	\centering
	\includegraphics[width=\linewidth]{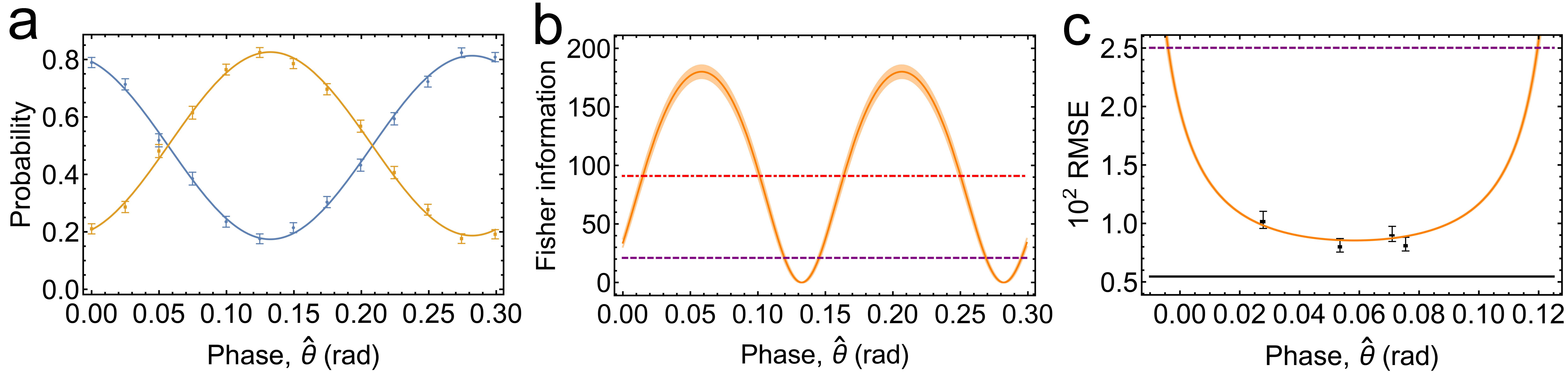}
	\caption{\textbf{Experimental results for the combined strategy.} \textbf{a}, The average outcome probability in the measurement basis $\sigma_x^{\otimes6}$ for six-photon to the probe state $|\phi_{\text{MePc}}\rangle$. The blue (orange) line represents the average outcome probability $P_{+}^{123456}$ ($P_{-}^{123456}$). \textbf{b}, The FI per trial, fitted from $P^{123456}$ for MePc. The red dot-dashed line is the theoretical limit of FI for MsPc. The purple dotted line is the theoretical limit of FI for the SNL. \textbf{c}, Observed phase estimation and its uncertainty. The black solid line is the theoretical limit of RMSE for MePc. The purple dashed line is the theoretical limit of RMSE for the SNL. The shaded areas correspond to the 90$\%$ confidence region, which are obtained from the uncertainty in the fitting parameters. The square data points are calculated from 7000 detection events. The error bars are discussed in the Methods.}
	\label{fig:randomness3}
\end{figure}

Next, we demonstrate the combined strategy with $\theta_1$ ramping continuously from $ 0 $ to $ 2\pi $. The experimental data are fitted to the function $\ P_{\pm}=\frac{1\pm V_\pm \cos(21\hat{\theta})}{2}$, where $V_{+}=0.647$ and $V_{-}=0.631$ denote the fringe visibilities in the eigenvectors $|\pm1\rangle$ of measurement basis ${\sigma_x}^{\otimes6}$. As shown in Fig.~\ref{fig:randomness3}a,b, we fit the observed average outcome probability values and calculate the FI according to Eq.~\eqref{eq:FI}. the maximum value of FI is about 180 at $\hat{\theta}=\frac{\pi}{21}$, which represents a 4.7 dB reduction compared with the SNL of MsPc.

Finally, we consider the range where we expect to beat the theoretical limit based on the probe state MePc. We take around 70 measurements to obtain the probabilities of the measurement outcomes. The estimator $\hat{\theta}=\sum_{k=1}^{6}{k \theta_k}/21$ is obtained using the maximum likelihood estimation, which maximizes the posterior probability based on the obtained data. To experimentally obtain the statistics of $\hat{\theta}$, we repeat the process 100 times to get the distribution of $\hat{\theta}$, from which the standard deviation of the estimator $\delta \hat{\theta}$ is obtained. As shown in Fig.~\ref{fig:randomness3}c, the experimental precision (black dots) saturates the theoretical optimum value.

\emph{Discussion and conclusion. ---}
Our experiment uses post-selection which does not include the experimental imperfections of probabilistic generation of photons from SPDC and the photon loss. The post selection is a standard technique in almost all (except for ref.\cite{slussarenko2017unconditional}) previous quantum metrology experiments\cite{walther2004broglie,mitchell2004super,nagata2007beating,resch2007time,higgins2007entanglement,gao2010experimental,bell2013multicolor}. In future, with the improvement of collection and detection efficiency\cite{slussarenko2017unconditional}, our set-up can be directly extended to the demonstration of unconditional violation of the SNL for multi parameters. Also, for the combined strategy, we assume that the samples have no absorptions and the samples' phases are uniformly distributed. However, these assumptions do not have influence on the proof-of-concept verification of the super-resolution effect.

Overall, we have demonstrated three types of strategies for distributed quantum metrology, by observing the visibility and FI of phase super-resolution. First, we demonstrate the estimation of individual parameters in three modes. All experimental fringes shown in Fig~\ref{fig:randomness1} present high visibility that is sufficient to beat the SNL. Second, by using a tunable interferometer, we estimate an averaged phase shift across three modes. The visibility of (MePe, MePs, MsPe) shown in Fig.~\ref{fig:randomness2} clearly demonstrates that MePe is the optimal choice with the highest sensitivity to estimate the averaged phase shift. The maximum value of FI is about 20.825, which beats all the theoretical bounds for MePs, MsPe and MsPs (Methods). Third, by interacting the photon through the samples multiple times in each mode, we demonstrate the combined strategy with parallel entanglement across six modes and photon passes up to $N=21$. Our results may open a new window for exploring the advanced features of entanglement and coherence in a quantum network for distributed quantum phase estimation, which may find quantum enhancements for sensing applications.

\section*{Data availability}
The data that support the plots within this paper and other findings of this study are available from the corresponding authors upon reasonable request.

\section*{Code availability}
The code that support the plots within this paper and other findings of this study are available from the corresponding authors upon reasonable request.

\section*{Reference}

\bibliographystyle{naturemag}
\bibliography{DQPEE}

\section*{Acknowledgements}
This work was supported by the National Key Research and Development (R\&D) Plan of China (under Grants No. 2018YFB0504300 and No. 2018YFA0306501), the National Natural Science Foundation of China (under Grants No. 11425417, No. 61771443 and U1738140), the Shanghai Municipal Science and Technology Major Project (Grant No. 2019SHZDZX01), the Anhui Initiative in Quantum Information Technologies and the Chinese Academy of Sciences.

\section*{Author contributions}
Z.-D.L., F.X., Y.-A.C. and J.-W.P. conceived and designed the experiments. Z.-D.L., F.X. and Y.-A.C. designed and characterized the multiphoton optical circuits. Z.-D.L., R.Z., X.-F.Y., L.-Z.L., Y.H., Y.-Q.F and Y.-Y.F carried out the experiments. Z.-D.L, R.Z., F.X. and Y.-A.C. analysed the data. Z.-D.L., X.F., Y.-A.C. and J.-W.P. wrote the manuscript, with input from all authors. F.X., Y.-A.C. and J.-W.P. supervised the project.

\section*{Competing financial interests}
The authors declare no competing financial or non-financial interests.

\section*{Additional information}
Correspondence and requests for materials should be addressed to F.X., Y.-A.C or J.-W.P.

\clearpage

\clearpage
\noindent \bm{{\Large Method}}

\section*{Sensitivity evaluation}

In our experiment, we assume that the form of objective function is $\hat{\theta} = \bm{\upalpha^\text{T}}\bm{\uptheta}$, where ${\bm{\uptheta}} = ({\theta}_1,\dots,{\theta}_M)$ and $ \bm{\upalpha}=(\alpha_1,\dots,\alpha_M)$ denote, respectively, the vector of phase shift and the normalization coefficients with $\sum_k \alpha_k=1$.
The Hamiltonians are set to $H_k=\sigma_z/2$ for mode $k =1,\dots,M$, where $\sigma_z$ is the pauli matrix. The unitary operator of mode $k$ can be expressed as
\begin{equation} \label{Uk}
U_{k} =
\begin{pmatrix}
e^{-i\theta_k/2} & 0 \\
0 & e^{i\theta_k/2}
\end{pmatrix}.
\end{equation}
According to the given evolution, we can determine the sensitivity for different estimation strategies.

(1) Let us start with the analysis of the parallel strategy. To obtain the optimal sensitivity for modes entangled and particles entangled (MePe), we will first consider the probe state that contain entanglement among all of the $N$ photons, that is, the Greenberger-Horne-Zeilinger (GHZ) state:
\begin{equation}
\begin{aligned}
&|\phi_{\text{MePe}}\rangle_{\rm ini}=\frac{1}{\sqrt{2}} \left( \Sigma_{k=1}^{M} |H\rangle^{\otimes N_k}+\Sigma_{k=1}^{M} |V\rangle^{\otimes N_k} \right ),
\end{aligned}
\end{equation}
where $N_k=\alpha_k N$ denotes the number of photons in mode $k =1,\dots,M$ and the total number of photons is $N = \sum_k N_k$. The probe state after the evolution as described in Eq.~\eqref{Uk} is of the form
\begin{equation}
\begin{aligned}
&|\phi_{\text{MePe}}\rangle_{\rm evo}=\frac{1}{\sqrt{2}} \left( \bigotimes_{k=1}^{M} |H\rangle^{\otimes N_k}+ e^{i\sum_{k=1}^{M} N_k \theta_k } \bigotimes_{k=1}^{M} |V\rangle^{\otimes N_k} \right ).
\end{aligned}
\end{equation}
The projective measurements on the probe state are performed in the $ \sigma_x $ basis, which can achieve the maximum visibility for interference fringe\cite{resch2007time,gao2010experimental}. In this setting, the outcome probability in the eigenvectors $|\pm1\rangle$ are
\begin{equation} \label{P}
P^{\rm MePe}_{\pm1} =\frac{1\pm V_{\pm} \cos( \Sigma_{k=1}^{M} n_k \theta_k )}{2}=\frac{1\pm V_{\pm} \cos( N \hat{\theta})}{2},
\end{equation}
where $V_\pm$ denotes the fringe visibilities in the eigenvectors $|\pm1\rangle$ of measurement basis ${\sigma_x}^{\otimes6}$. Following the above expression, the FI of $|\phi_{\rm MePe}\rangle$ can be calculated as
\begin{equation} \label{FI}
F_{\rm MePe}(\hat{\theta})=\frac{V^2 N^2 \sin^2(N \hat{\theta})}{1-V^2\cos^2(N \hat{\theta})}.
\end{equation}
It is easy to see that the Heisenberg limit $\delta \hat{\theta}=1/\sqrt{F_{\rm MePe}}=1/N$ can be achieved when the noise is free $(V=1)$.

We then consider the sensitivity for modes entangled and particles separated (MePs). For the purpose of our study, the equal weight linear function is considered and the probe state reach the optimal sensitivity can be written as
\begin{equation}
\begin{aligned}
|\phi_{\text{MePs}}\rangle_{\rm ini}=\frac{1}{2}\left (\Sigma_{k=1}^{M}|H\rangle_k+\Sigma_{k=1}^{M}|V\rangle_k\right )^{\otimes \frac{N}{M}},
\end{aligned}
\end{equation}
where we assume $N/M$ is an integer. After the evolution, this probe state becomes to
\begin{equation}
\begin{aligned}
|\phi_{\text{MePs}}\rangle_{\rm evo}=\frac{1}{2}\left (\otimes_{k=1}^{M}|H\rangle_k+ e^{i\sum_{k=1}^{M} \theta_k } \otimes_{k=1}^{M}|V\rangle_k\right )^{\otimes \frac{N}{M}}.
\end{aligned}
\end{equation}
Since the objective function is $\hat{\theta} = \bm{\alpha^T}\bm{\theta}$, and  ${|\phi_{\text{MePs}}\rangle}_{\rm evo}$ is a product state of $N/M$ identical $M$-mode entanglement states, the theoretical limit for MePs is the sum of the FI of these $N/M$ states, that is, $\delta \hat{\theta}=1/\sqrt{F_{\rm MePs}}=1/\sqrt{MN}$.

Indeed, the protocol for modes separated and particles entangled (MsPe) can be viewed as estimating the parameters separately in each mode\cite{gessner2018sensitivity}. The sensitivity is converged to $\delta \hat{\theta}=1/\sqrt{F_{\rm MsPe}}=1/\left (\sum_{k=1}^{M} N_k^2\right )^{\frac{1}{2}}$, which is equal to the sum of of single-parameter sensitivities.

In our experiment, we set the number of photons $N=6$ and the number of modes $M=3$. Therefore, according to above conclusions, the FI are $F_{\rm MePe}=36$, $F_{\rm MePs}=18$ and $F_{\rm MsPe}=12$ when the noise is free $(V=1)$.

(2) In the combined strategy, we utilize the coherence rather than the particle entanglement in each mode. In this case, the essential feature is that the phase shift is being interacted coherently over many passes of the unitary evolution. This process can be described as follows
\begin{equation}
\begin{aligned}
&|\phi_{\text{MePc}}\rangle=\frac{1}{\sqrt{2}}\left (\Sigma_{k=1}^{M}|H\rangle_k+\Sigma_{k=1}^{M}|V\rangle_k\right ) &\\
\xrightarrow[]{{\bm{\uptheta}} }& \frac{1}{\sqrt{2}}\left (\Sigma_{k=1}^{M}|H\rangle_k+e^{i\sum_{k=1}^{M} \theta_k }\Sigma_{k=1}^{M}|V\rangle_k\right ),
\end{aligned}
\end{equation}

where $n_k=\alpha_k n$ denotes the number of interactions in mode $k =1,\dots,M$ and the total number of interactions is $n = \sum_k n_k$. When the noise is free $(V=1)$, the Heisenberg limit can be achieved for MePc, that is, $\delta \hat{\theta}=1/\sqrt{F}=1/n$. The sensitivity for mode separated and particle conherent (MsPc) is converged to $\delta \hat{\theta}=1/\sqrt{F_{\rm MsPe}}=1/\left (\Sigma_{k=1}^{M} n_k^2\right )^{\frac{1}{2}}$. In our experiment, we set the number of interactions $n=\Sigma_{k=1}^{6} k=21$ and the number of modes $M=6$, and thus the FI are $F_{\rm MePc}=441$, $F_{\rm MsPc}=91$.

\section*{Error analysis}

To obtain the standard deviation of the value of phase shifter, we take $k$ measurement sets, and each set contains around $s$ coincidence events. In our experiment, around 7000 coincidence events are measured and divided into 100 groups for each phase shifter. By using maximum likelihood method, the standard deviation $\delta \hat{\theta}$ is then obtained from the outcome probability, which are calculated from these coincidence events. The error for this experimentally obtained $\delta \hat{\theta}$ is well approximated by $\delta \hat{\theta}=\delta \hat{\theta}/\sqrt{2(s-1)}$ \cite{hou2019control}.

\section*{Extended Data Figure 1}
\begin{figure*}[htbp]
	\centering
	\includegraphics[width=0.6 \linewidth]{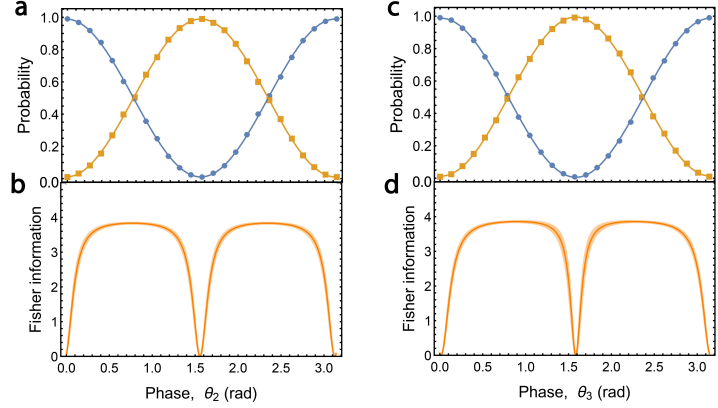}
	\caption{\textbf{Experimental results for the parallel strategies of MsPe for mode 2 and mode 3.} \textbf{a,c,} The average outcome probability in the measurement basis $\sigma_x^{\otimes2}$ for two-photon entangled states ($2^{\prime}5^{\prime}$ and $4^{\prime}6$ respectively). Blue (orange) lines represent the average outcome probability $P_{+}^{2^{\prime}5^{\prime}}$ and $P_{+}^{4^{\prime}6}$ ($P_{-}^{2^{\prime}5^{\prime}}$ and $P_{-}^{4^{\prime}6}$) for mode 2 and mode 3. \textbf{b,d,} The fisher information per trial, fitted from $P^{2^{\prime}5^{\prime}}$ and $P^{4^{\prime}6}$ for MsPs, respectively. The shaded areas correspond to the $90\%$ confidence region, derived from uncertainty in the fitting parameters. Error bars are calculated from measurement statistics and too small to be visible. Red dot-dashed line: the theoretical limit for MePs. Blue dashed line: the theoretical limit of MsPe. Black dotted line: the theoretical value of the SNL for MsPs.}
\end{figure*}

\end{document}